# NON-LINEAR TECHNIQUES FOR INCREASING HARVESTING ENERGY FROM PIEZOELECTRIC AND ELECTROMAGNETIC MICRO-POWER-GENERATORS

*Yasser AMMAR, Skandar BASROUR*

*TIMA, 46 avenue Félix Viallet, 38031 Grenoble Cedex, France.*

## ABSTRACT

Non-linear techniques are used to optimize the harvested energy from piezoelectric and electromagnetic generators. This paper introduces an analytical study for the voltage amplification obtained from these techniques. The analytical study is experimentally validated using a macro model of piezoelectric generator. Moreover, the integration influences on these techniques is studied. Through all the obtained results, a suitable structure for autonomous microsystems is proposed.

## 1. INTRODUCTION

Autonomous Microsystems are new category of systems. They convert available environmental energy to useful electrical energy. The environmental energy can be heat, light or mechanical vibrations. An interesting domain for using autonomous Microsystems is human applications, where an important amount of free clean and durable energy is available through body movements. Moreover, autonomous Microsystems can be used in industrial applications for monitoring different physical parameters.

During the last decade, many researches have been performed in order to harvest energy from mechanical vibrations. Electromagnetic generators are proposed by Amirtharajah et al. [1], as well as by Williams et al.[2]. Electrostatic generators are other types of generators that use vibrations to generate electrical energy. Examples of these generators are proposed by Mitcheson et al. [3], and Despesse et al. [4]. Finally piezoelectricity is another way to convert mechanical energy into electricity [6].

In this paper, we will propose and discuss different techniques in order to maximize the energy harvested from piezoelectric and electromagnetic micro power generators.

Reducing the dimensions of macro generators leads to low energy and low output voltage. So the design of energy harvesting circuits will be based on ultra low power consumption and high efficiency architectures. Up to now, the proposed approaches consist of adding conventional AC/DC converter after the micro-power-generators. The AC/DC converter is followed by an adaptive DC/DC converter, which is used for impedance matching [5]. Due to the low voltages obtained on micro-power generators new approaches are necessary.

Guyomar et al. have proposed a non-linear technique for piezoelectric macro generators called Synchronized Switch Harvesting on Inductor (SSHI) [7]. The SSHI needs inductors with high quality factors, which are difficult to obtain with integrated micromachining techniques. We will propose a novel method called Synchronized Switch Harvesting on Capacitor (SSHC). The SSHC uses capacitors instead of inductors. Moreover, this method is more tolerant regarding to the time of switching. The SSHC is the dual approach of the SSHI for electromagnetic power generators.

Starting from analytical model for piezoelectric scavengers, the expressions of the voltage gain for the two techniques have been deduced. With typical values for inductors, the gain can reach ten times the initial output voltage provided by the micro-power generator.

This paper starts with a brief introduction of energy harvesting circuit used in autonomous microsystems. In the third section, the model of piezoelectric generators is presented. The fourth section contains the principles of non-linear techniques. The analytical and the experimental studies of voltage amplification are introduced in the fifth and sixth sections. Then an explanation of the integration influence on these techniques is studied. As a result a solution based on these techniques for autonomous microsystems is presented.

## 2. ENERGY HARVESTING CIRCUIT

An autonomous micro system uses environment energy as power source. This study will be restricted on systems powered by vibrations. The schematic of harvesting energy is represented in Figure 1. In this figure, the signal given by the power generator is rectified using an AC/DC converter. Then a DC/DC converter is used to match the output of AC/DC converter with the load.

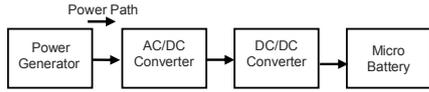

Figure 1 : The basic structure of harvesting energy.

The generated power as well as the voltage in the case of micro power generators is very low. Hence, it is necessary to optimize the harvested energy from the micro power generator. The non-linear techniques are used to perform this optimization as shown in Figure 2.

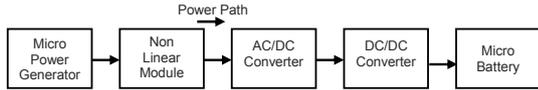

Figure 2 : The improved structure of harvesting energy.

### 3. MODELING

The Figure 3 shows a basic example of the piezoelectric generators. It consists of a bimorph piezoelectric beam excited in the mode 31. The piezoelectric equations are given in (1):

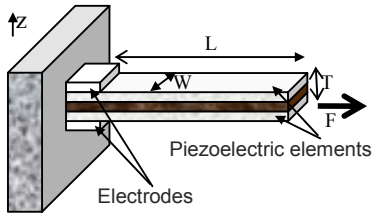

Figure 3 : The basic structure of a piezo-generator.

$$S_i = s_{11}T_i + d_{31}E_3$$
$$D_3 = d_{31}T_i + \varepsilon_{33}E_3 \qquad (1)$$

where

$S_i$      is the strain in x direction.

$T_i$      is the stress (N/m$^2$).

$E_3$      is the electric field in z direction (V/m).

$D_3$      is charge density (C/m$^2$).

$s_{11}$      is compliance coefficient (m$^2$/N).

$d_{31}$      is piezoelectric coupling coefficient (C/N).

These equations lead to the equivalent electrical model for the piezoelectric generator [8]. Figure 4 shows the electrical model, which consists of a current generator in parallel with a capacitor. A resistor $R_p$ is added to the model to represent the dielectric losses.

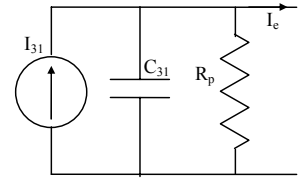

Figure 4 : The equivalent circuit of piezoelectric generator.

Where

$I_{31} = NC_m \dfrac{dF}{dt}$    is the equivalent current source.

$N = \dfrac{d_{31}L_{22}}{s_{11}}$    is the coupling coefficient.

$C_{31} = N^2 C_m + C_0$

$C_m = \dfrac{s_{11}L_1}{A_1}$    is the equivalent capacitor.

$C_0 = \dfrac{A_3 \varepsilon_{33}}{L_3}$

### 4. NON-LINEAR TECHNIQUES

#### SSHI METHOD

This method consists of switching an inductor in parallel to the piezoelectric generator as seen in Figure 5. The inductor is switched when the displacement is maximal. The period of switch T is equal to one-half of the period of the oscillator composed of the inductor L and the capacitor $C_p$ (the piezo-capacitor).

$$T = \pi \sqrt{LC_p} \qquad (3)$$

The switching of the inductor causes the inversion of the piezoelectric generator voltage. However, the inversion is not perfect because of losses due to $R_{load}$. Figure 6 represents the inversion phenomena.

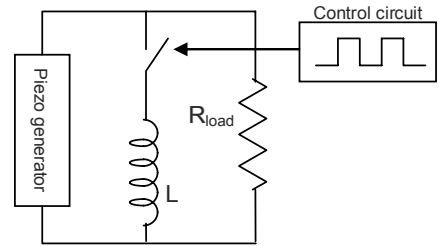

Figure 5 : SSHI method (schematic).

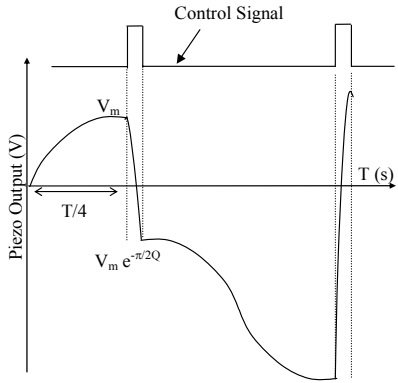

Figure 6 : SSHI method (waveform).

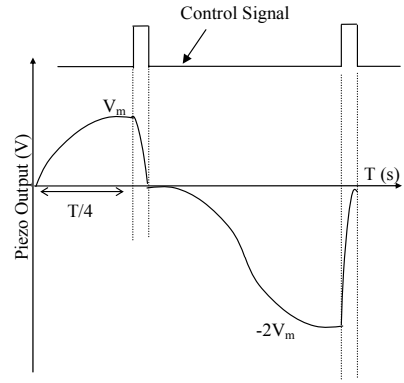

Figure 7 : SSHC method (waveform).

The following parameters are considered:

$$\omega_0 = \sqrt{\frac{1}{LC_p}}$$

$$2\lambda\omega_0 = \frac{1}{R_{load}C_p}$$

$$\lambda = \frac{1}{2Q} \qquad (4)$$

$$\alpha_{inv} = e^{-\lambda\pi}$$

$$\alpha_{rc} = e^{\frac{-T}{2R_{load}C_p}}$$

$$\alpha_a = \frac{1+\alpha_{rc}}{1-\alpha_{inv}\alpha_{rc}}$$

Equation (5) gives the analytical value of voltage amplification.

$$A = 1 + (1 - \alpha_a\alpha_{inv})\alpha_{rc} \qquad (5)$$

The value of the amplification depends on the value of $R_{load}$, $C_p$ and $\omega_0$.

### SSHC METHOD

SSHC method is based on switching a capacitor in parallel to the piezoelectric element. The value of the added capacitor is greater or equal to ten times the value of the piezoelectric capacitor. The moments of switching are synchronized with the maximum and minimum voltage.

The SSHC technique is presented in Figure 7.

Equation (6) gives the analytical value of voltage amplification.

$$A = 1 + e^{\frac{-T}{R_{load}C_p}} \qquad (6)$$

The maximum value of amplification in SSHC method equals two.

## 5. EXPERIMENTAL RESULTS

### SSHI METHOD

The bimorph represented in Figure 1 is used as a macro generator (L=35mm, W=12.5mm, T=0.45mm with a mass attached to its end equals 2.5g). It is connected to a shaker, which generates sinusoidal excitation. The bimorph is excited by a sinusoidal signal with a frequency equals to 85Hz, which equals the frequency of resonance for this structure. The used acceleration equals to 1ms$^{-2}$. The method of SSHI has been realised by using an inductor of 22mH and a switch composed of two NMOS transistors. An FPGA circuit is used to generate synchronised pulses at the gates of the transistors. The Figure 8 represents the structure of experimental circuit.

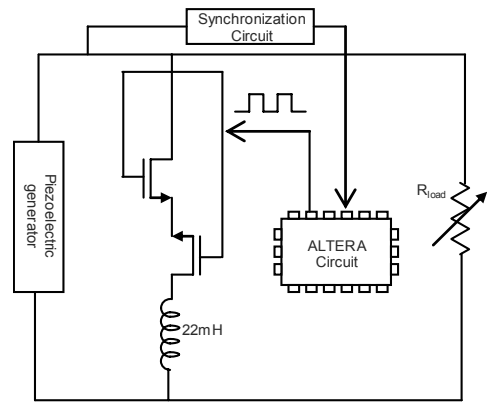

Figure 8 : The experimental realization of SSHI.

Figure 9 represents good agreement between analytical and experimental study.

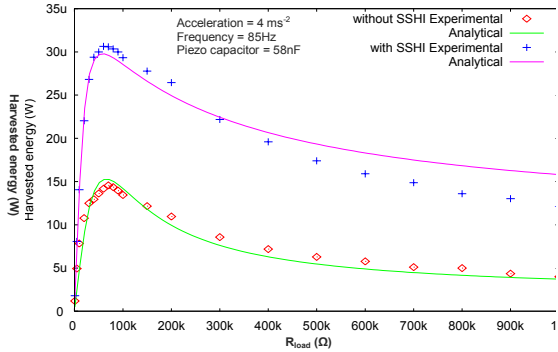

Figure 9 : The comparison between experimental results and analytical study.

### SSHC METHOD

SSHC technique is realised by using the same circuit of SSHI shown in Figure 8. A capacitor of 1µF substitutes the inductor. Figure 10 represents the experimental results using the SSHC technique. The amplification of the voltage equals 1.76. The analytical value of amplification equals 1.66.

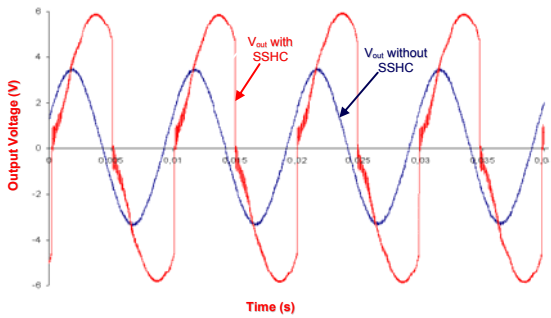

Figure 10 : The experimental results for SSHC technique.

## 6. INTEGRATION

### INTEGRATION OF SSHI

The integration of SSHI means the integration of the inductor, the switch and the control circuit.

Equivalent and simplified models for integrated inductor are presented in Figure 11.

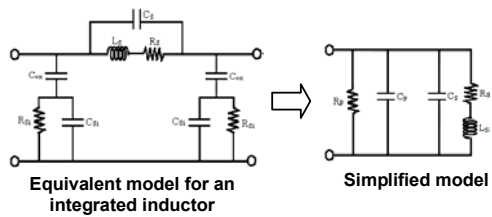

Figure 11 : The equivalent circuits for integrated inductor.

Comparing to piezoelectric generator components, $R_p$, $C_p$ and $C_s$ can be neglected; therefore, the remaining element is the serial resistance $R_s$. Similarly, a resistance can represent the equivalent circuit of MOS switch (in the on state). The equivalent circuit of the piezoelectric generator with an integrated switch and inductor is shown in Figure 12.

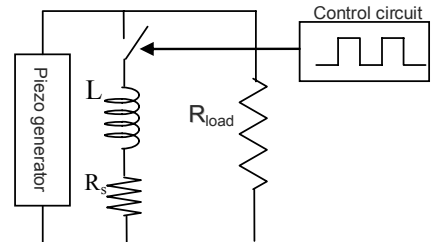

Figure 12 : The model of a piezoelectric generator with an integrated inductor and switch.

In this case, equations (6) give the changes introduced on equations (4). All other equations used to calculate the amplification are still valid:

$$\omega_0 = \sqrt{\frac{1 + \dfrac{R_s}{R_{load}}}{LC_p}}$$

$$2\lambda\omega_0 = \frac{1}{R_{load}C_p} + \frac{R_s}{L} \qquad (6)$$

$$\lambda = \frac{1}{2Q}$$

Figure 13 represents the estimated value of amplification versus the load value for different integrated inductors. This figure shows that the amplification for an inductor of 1.07nH fabricated in AMS0.35µm technology equals to 2.2.

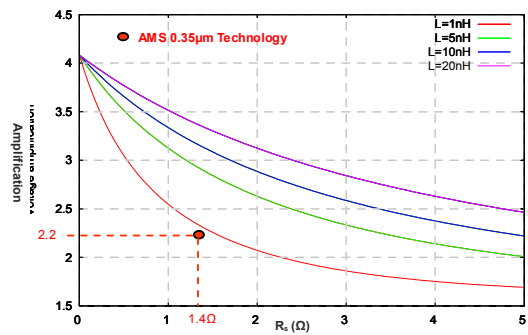

Figure 13 : The analytical results of amplification for integrated SSHI.

As a result, the integrated SSHI method is efficient only for integrated inductors with small value of serial resistance. In other words, it is efficient for inductors with high value of quality factor.

An auto control circuit for SSHI is proposed; this circuit contains a comparator, two inverters and six DFF. The design of comparator is optimized in order to minimize the power consumption. The other components are used from the library of AMS 0.35µm. The result of simulation for the auto synchronization is presented in Figure 14.

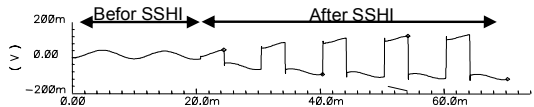

Figure 14 : The simulation result for SSHI.

### *INTEGRATION OF SSHC*

The integration of capacitors has not the same influence as the integration of inductors in SSHI technique. The integration will add a serial resistor to the structure. This resistor will increase the time of discharging the piezoelectric capacitor. Therefore, the time of switch-on should be increased in order to discharge this capacitor.

To complete the path of harvesting energy, a structure of ultra low power AC/DC is proposed. This structure can rectify signals of amplitude less than the threshold voltage of diodes. The consumption of this converter is about 60nW.

The estimated consumption for SSHC circuit control is 300nW. At the output of AC/DC converter, we can harvest 4p-300nW as seen in Figure 15.

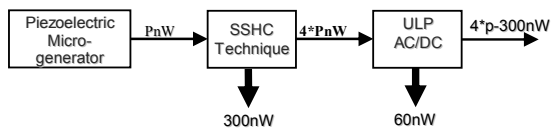

Figure 15 : The estimation of power consumption for harvesting circuit.

This circuit with the block DC/DC are in the phase of optimizing in order to obtain the maximum efficiency.

### 7. PERSPECTIVES

The model of an electromagnetic generator is presented in Figure 16. This model consists of a voltage generator in series with an inductor $L_m$ a resistor $R_m$. It is clear that the electromagnetic generator is the dual of piezoelectric generator. Hence, it seems to be evident to apply a similar study made for the piezoelectric generators. In the case of electromagnetic generators, the method SSHC will give amplification similar to the application of SSHI method for piezoelectric generators. This approach is in the phase of validation experimentally.

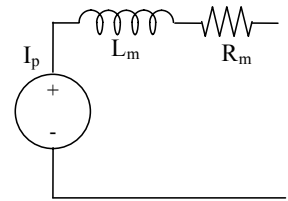

Figure 16 : The model of electromagnetic generator.

### 8. CONCLUSION

The two techniques for optimizing harvesting energy from piezoelectric generators are presented in this paper. Moreover, an analytical study for amplification was compared to the experimental results. It is shown that the integration of SSHI is critical because of the low quality factor of integrated inductors. Therefore, the technique of SSHC seems to be more convenient in the case of integration. In addition, this technique can be used for electromagnetic generators.